\documentclass[Physsubmission, Phys]{SciPost}

\hypersetup{
    colorlinks,
    linkcolor={red!50!black},
    citecolor={blue!50!black},
    urlcolor={blue!80!black}
}

\usepackage[bitstream-charter]{mathdesign}
\DeclareSymbolFont{usualmathcal}{OMS}{cmsy}{m}{n}
\DeclareSymbolFontAlphabet{\mathcal}{usualmathcal}

\newcommand{\bp}{\mbox{\boldmath $p$}}
\newcommand{\bq}{\mbox{\boldmath $q$}}

\newcommand{\bP}{\mbox{\boldmath $P$}}

\newcommand{\bk}{\mbox{\boldmath $k$}}

\newcommand{\ket}[1]{| {#1} \rangle}

\begin{document}

\begin{center}{\Large \textbf{
Prompt hadroproduction of C-even quarkonia \\ in the
light-front $k_T$ -factorization approach \\
}}\end{center}

\begin{center}
Wolfgang Sch\"afer\textsuperscript{1$\star$},
Izabela Babiarz\textsuperscript{1},
Roman Pasechnik \textsuperscript{2} and
Antoni Szczurek \textsuperscript{1}
\end{center}

\begin{center}
{\bf 1} Institute of Nuclear Physics Polish Academy of Sciences, Krak\'ow, Poland
\\
{\bf 2} Department of Astronomy and Theoretical Physics, Lund University, SE-223 62 Lund, Sweden
\\
* Wolfgang.Schafer@ifj.edu.pl
\end{center}

\begin{center}
\today
\end{center}


\definecolor{palegray}{gray}{0.95}
\begin{center}
\colorbox{palegray}{
  \begin{tabular}{rr}
  \begin{minipage}{0.1\textwidth}
    \includegraphics[width=22mm]{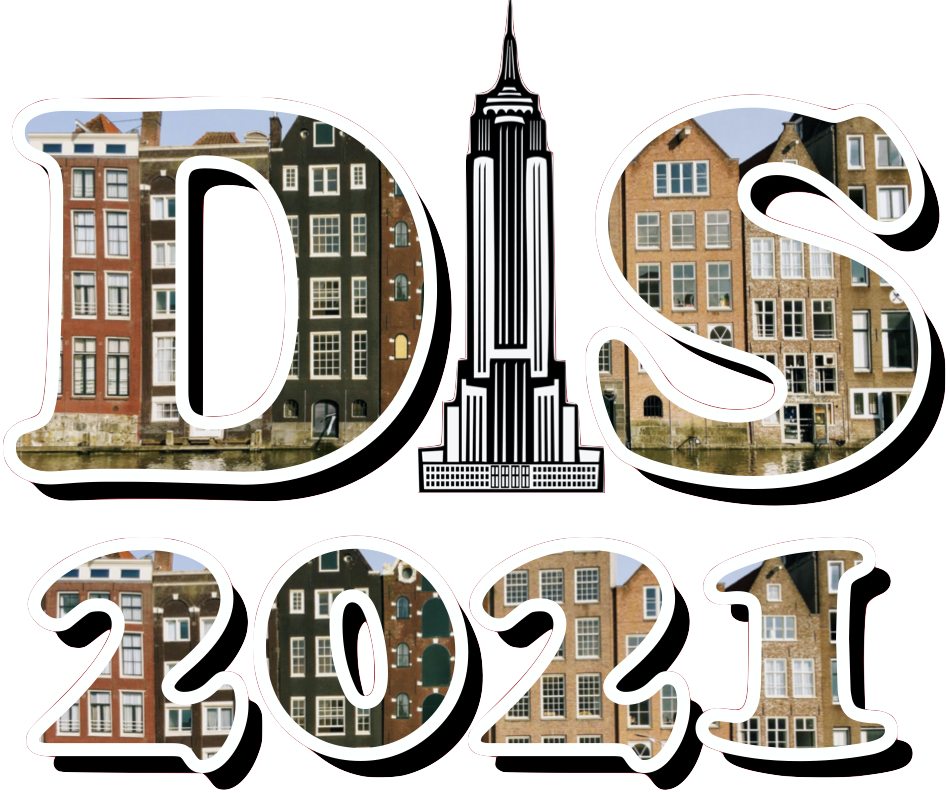}
  \end{minipage}
  &
  \begin{minipage}{0.75\textwidth}
    \begin{center}
    {\it Proceedings for the XXVIII International Workshop\\ on Deep-Inelastic Scattering and
Related Subjects,}\\
    {\it Stony Brook University, New York, USA, 12-16 April 2021} \\
    \doi{10.21468/SciPostPhysProc.?}\\
    \end{center}
  \end{minipage}
\end{tabular}
}
\end{center}

\section*{Abstract}
{\bf
We present a new approach for the prompt production of quarkonia which is based on the
$k_T$-factorization method. The production of even C-parity quarkonia proceeds via
the fusion of two (off-shell) gluons. Especially in the kinematics of the LHCb experiment
these processes are thus expected to be a sensitive probe of the small-$x$ gluon distribution.
We calculate the relevant off-shell matrix elements in terms
of the light-front wave functions of the quarkonium states. We present our results for scalar and pseudoscalar charmonia and discuss
photon transition form factors as well as cross sections for prompt hadroproduction.
  We compare our results for the $\eta_c$ to recent  LHCb data.
}


\section{Introduction}
\label{sec:intro}
The production of $C$-even quarkonia at high energies is a sort of ``Drell-Yan'' process for
gluons. Indeed, within the color-singlet approach the leading-order production mechanism
is the fusion of two gluons into the color-singlet heavy quark-antiquark $Q \bar Q$ pair. Provided the formation of the charmonium bound state is well understood, the process can be thought as a probe of the gluon densities. For a recent review of current challenges in quarkonium production, see \cite{Lansberg:2019adr}.
A typical hard scale for the process is the
transverse mass $m_T = \sqrt{\bp^2 + M^2}$, where $\bp$ is the transverse momentum, and $M$ the mass of the meson. In the high-energy regime, where we have $\Lambda_{\rm QCD} \ll m_T \ll \sqrt{s}$, transverse momenta of small-$x$ gluons can be taken into account in the $k_T$-factorization approach, which for heavy quark production has been initiated in \cite{Catani:1990eg,Levin:1991ry,Collins:1991ty}.
The crucial ingredient for the calculation is the off-shell matrix element (or impact factor) which describes the $g^* g^* \to \rm{Meson}$ subprocess.

In the color--singlet model, and in the lowest order of perturbation theory, the relevant off-shell matrix elements are directly propotional to $\gamma^* \gamma^*$ transition form factors of the meson. Here, we give a brief summary of our recent work \cite{Babiarz:2019sfa,Babiarz:2019mag,Babiarz:2020jkh}, where we discussed the cases of spinless, pseudoscalar and scalar, quarkonia.

\section{Brief overview of the formalism and some results}
\begin{figure}[!h]
        \centering
        a) \includegraphics[width=0.45\textwidth]{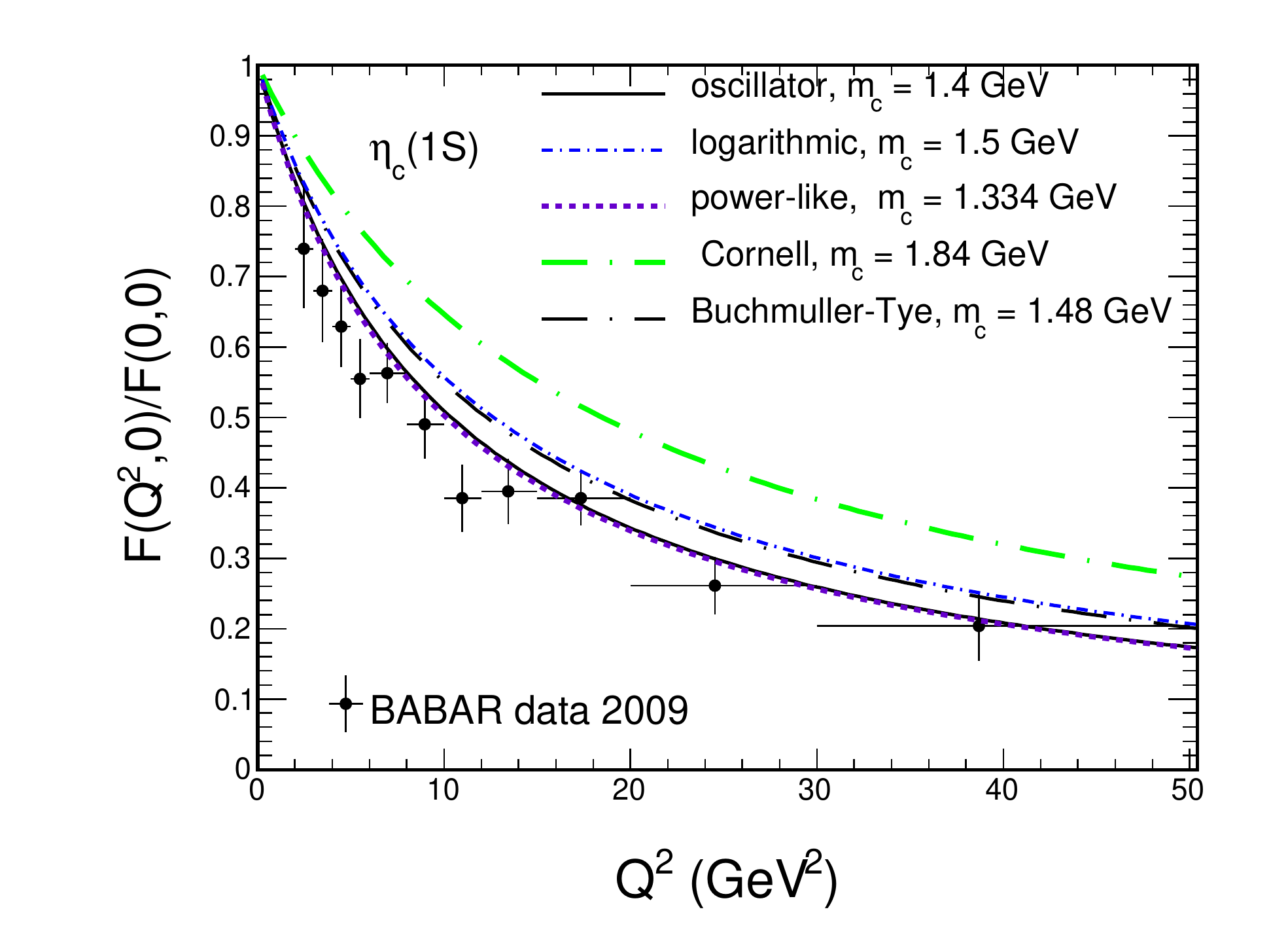}
        b) \includegraphics[width=0.45\textwidth]{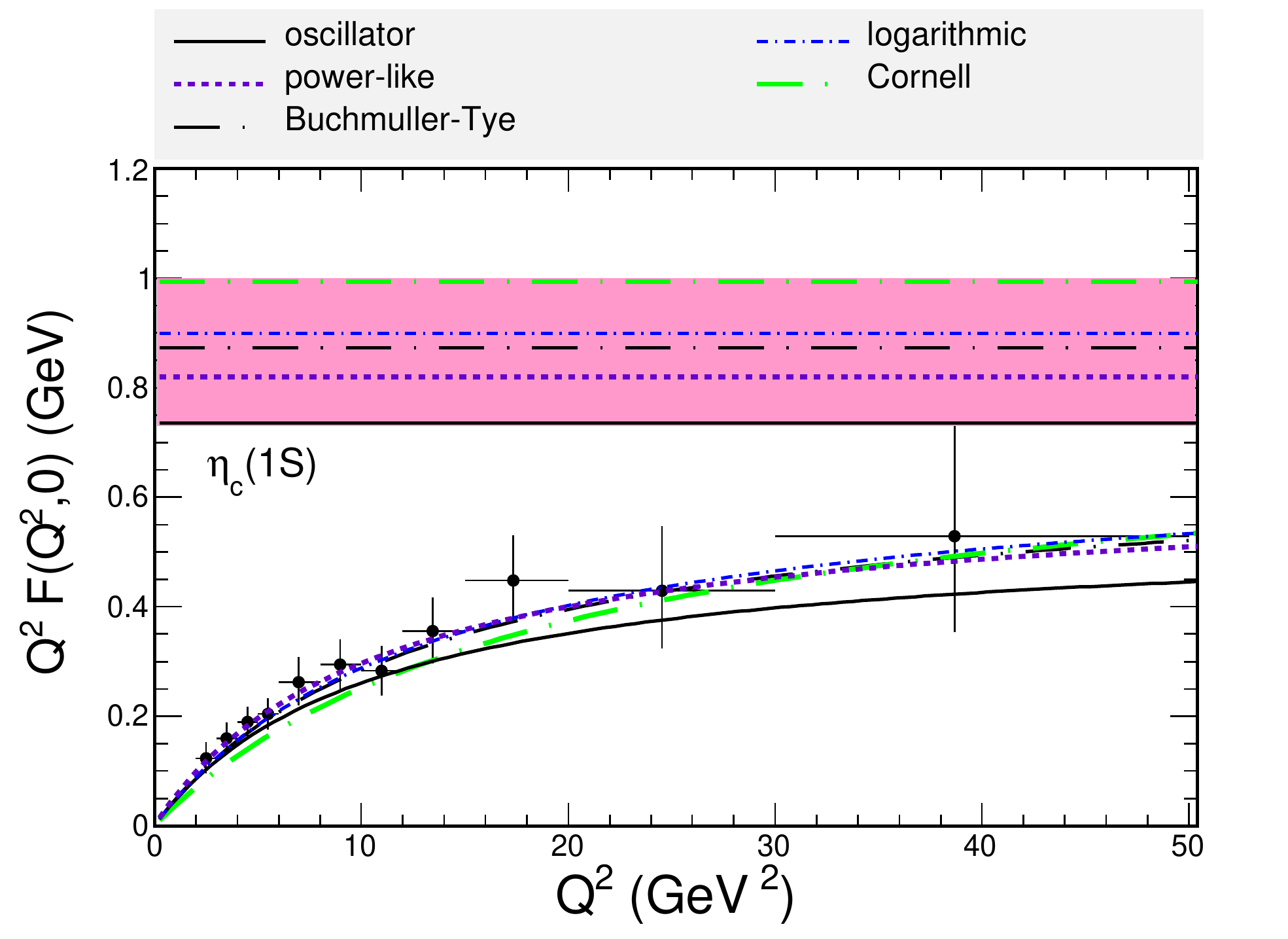}
        \caption{$\gamma^* \gamma$-transition form factor $F(Q^2,0)$ of $\eta_c$. In panel a) we show the transition form factor normalized to its value at the on-shell point. Theoretical results for a variety of potential model WFs are shown. The data are from Ref.~\cite{BaBar:2010siw}. In panel b) we show $Q^2 F(Q^2,0)$. The band shows the asymptotic limit for several potential models.}
        \label{fig:formfactors}
\end{figure}
We are interested in processes, where mesons are generally produced in highly boosted states. The light front approach allows us to define constituent structure of hadrons in a frame independent manner. 
We treat quarkonia as states dominated by the $Q \bar Q$ Fock component: 
\begin{eqnarray}
	\ket{{\rm Meson}; P_+, \bP}= \sum_{i,j,\lambda, \bar \lambda}
	{\delta^i_j \over \sqrt{N_c}} \, 
	\int {dz d^2\bk \over z(1-z) 16 \pi^3} \Psi_{\lambda \bar \lambda}(z,\bk)
	\ket{Q_{i \lambda}(z P_+ ,\bp_Q)
		\bar Q^j_{\bar \lambda}((1-z)P_+,\bp_{\bar Q})} + \dots \, .
	\end{eqnarray}
Here $\Psi_{\lambda \bar \lambda}(z,\bk)$ is the light-front (LF) wave function (WF), which depends on quark (light-front) momentum fraction $z$ and the relative momentum $\bk = (1-z) \bp_Q - z \bp_{\bar Q}$. For weakly bound, nearly non-relativistic systems, following Terentev, one can relate the light-front WF to the rest frame WF of a nonrelativistic potential model for quarkonia.
	\begin{eqnarray}
	\Psi_{\tau \bar \tau} (\vec k)
	&=& \underbrace{ \xi^{\tau \dagger}_Q \, \hat{\cal O} \, i \sigma_2 \, \xi^{\bar \tau *}_{\bar Q}}_{{\rm spin-orbit}} \, 
	\underbrace{{u_L(k) \over k}}_{\rm radial}{1 \over \sqrt{4 \pi}}; \, 
	\text{where}  \, \hat {\cal O} = \begin{cases}
	\mathbb{I}/\sqrt{2}  & \text{spin-singlet, $S=0,L=0$}.\\
	{\vec \sigma \cdot \vec k \over \sqrt{2} k} , & \text{spin-triplet, $S=1,L=1$}.
	\end{cases} 
	\label{eq:RF_WF}
	\end{eqnarray}

The relation between variables $z,\bk$ and the rest-frame (RF) relative momentum $k$ is simple, the nontrivial part is the relation between RF and LF spinors, given by the Melosh-transform: 
\begin{eqnarray}
	\xi_Q = R(z,\bk) \chi_Q , \, 	\xi^*_{\bar Q} = R^*(1-z,-\bk) \chi^*_{\bar Q} \, \, \text{with} \, \,  
	R(z, \bk) = { m_Q + z M - i \vec \sigma \cdot ( \vec n \times \bk) \over \sqrt{ (m_Q + z M)^2 + \bk^2  }} \, , \nonumber
	\end{eqnarray}
	so that
\begin{eqnarray}
	\hat {\cal O}' = R^\dagger(z,\bk) {\cal O} \, i \sigma_2
	R^*(1-z,-\bk) (i \sigma_2)^{-1} =  
	R^\dagger(z,\bk) \hat {\cal O} \, R(1-z,-\bk) \, ,
	\end{eqnarray}
which implies 
\begin{eqnarray}
	\Psi_{\tau \bar \tau} (\vec k) \to \Psi_{\lambda \bar \lambda} (z,\bk) = \chi_Q^{\lambda \dagger} {\cal O}' \, i \sigma_2 \chi^{\bar \lambda *}_{\bar Q} \, \cdot  \phi(z,\bk) \, . 
\end{eqnarray}

\begin{figure}[!h]
        \centering
        a) \includegraphics[width=0.4\textwidth]{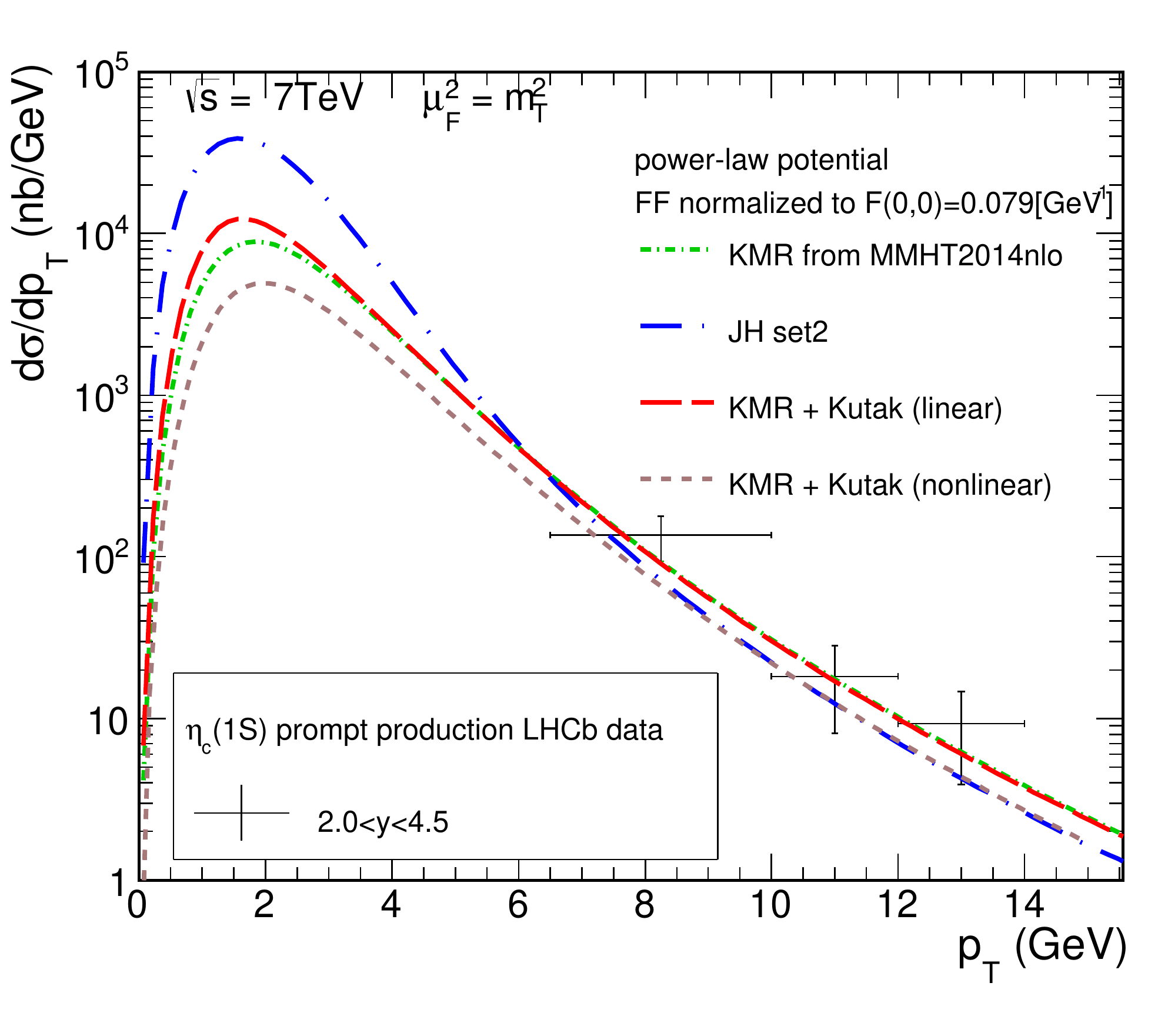}
        b) \includegraphics[width=0.4\textwidth]{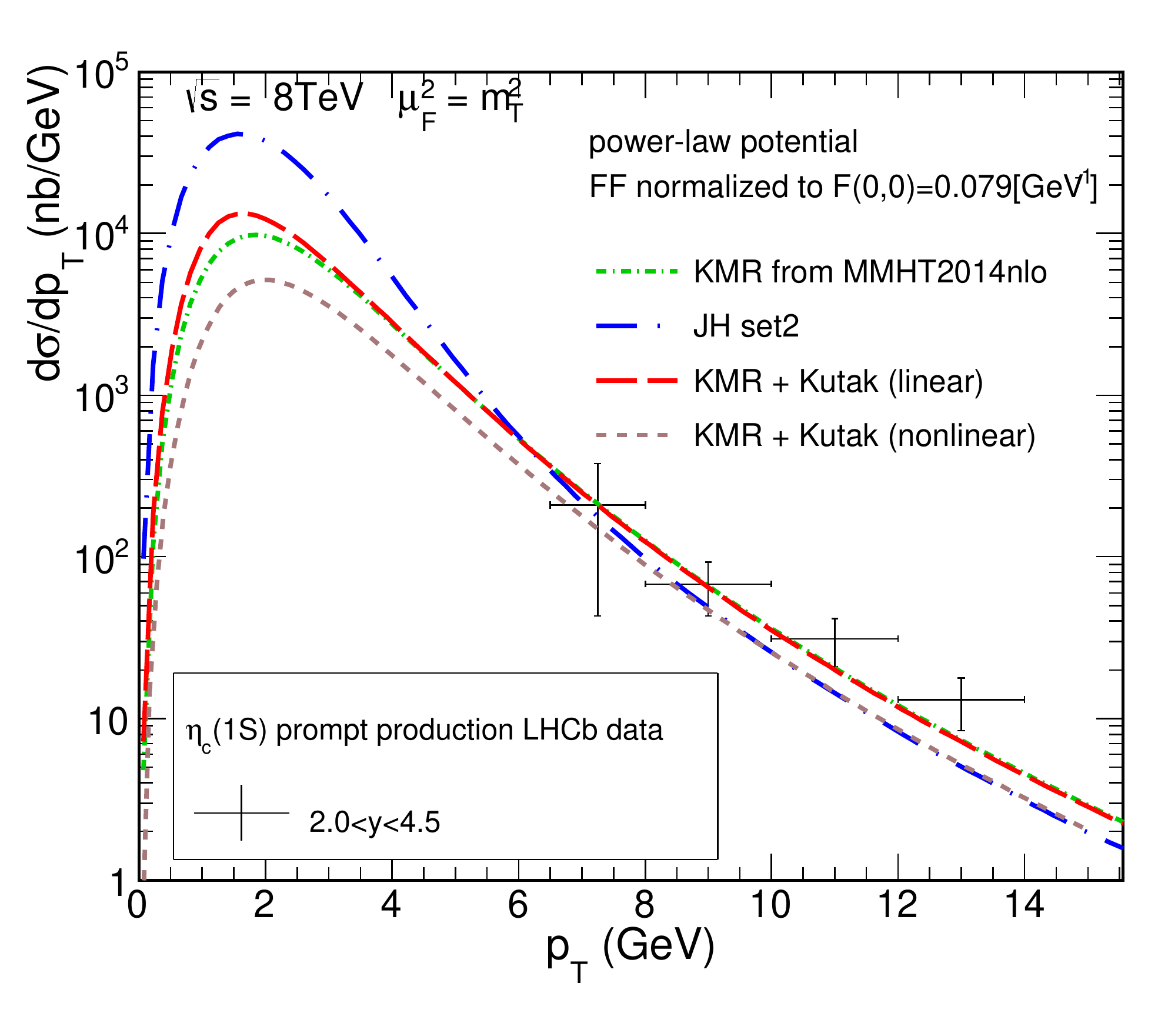} \\
        c) \includegraphics[width=0.4\textwidth]{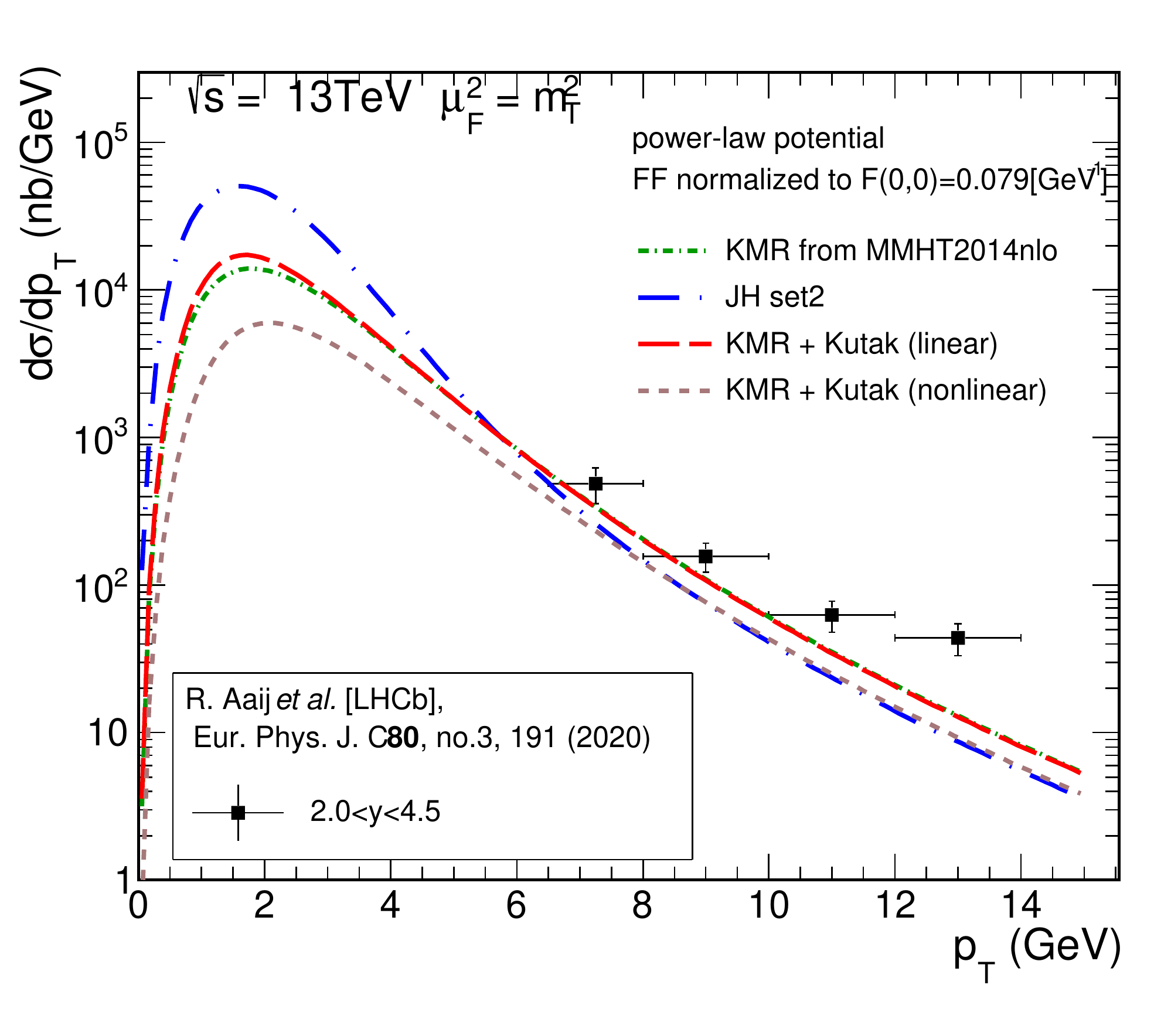}
        \caption{Prompt $\eta_c(1S)$ production:
        We show the differential cross section as a function of transverse momentum  at a) $\sqrt{s}= 7 \, \rm TeV$, b) $\sqrt{s}= 8 \, \rm TeV$, and c) 
        $\sqrt{s} = 13 \, \rm{TeV}$. 
        Data are from the LHCb Collaboration \cite{LHCb:2014oii}.
        Figs. a) and b) are found in \cite{Babiarz:2019mag}, while Fig. c) is an update showing the now published data of Ref. \cite{LHCb:2019zaj}.}
        \label{fig:dsig_dpt_etac1S}
\end{figure}

\begin{figure}[!h]
        \centering
        a) \includegraphics[width=0.4\textwidth]{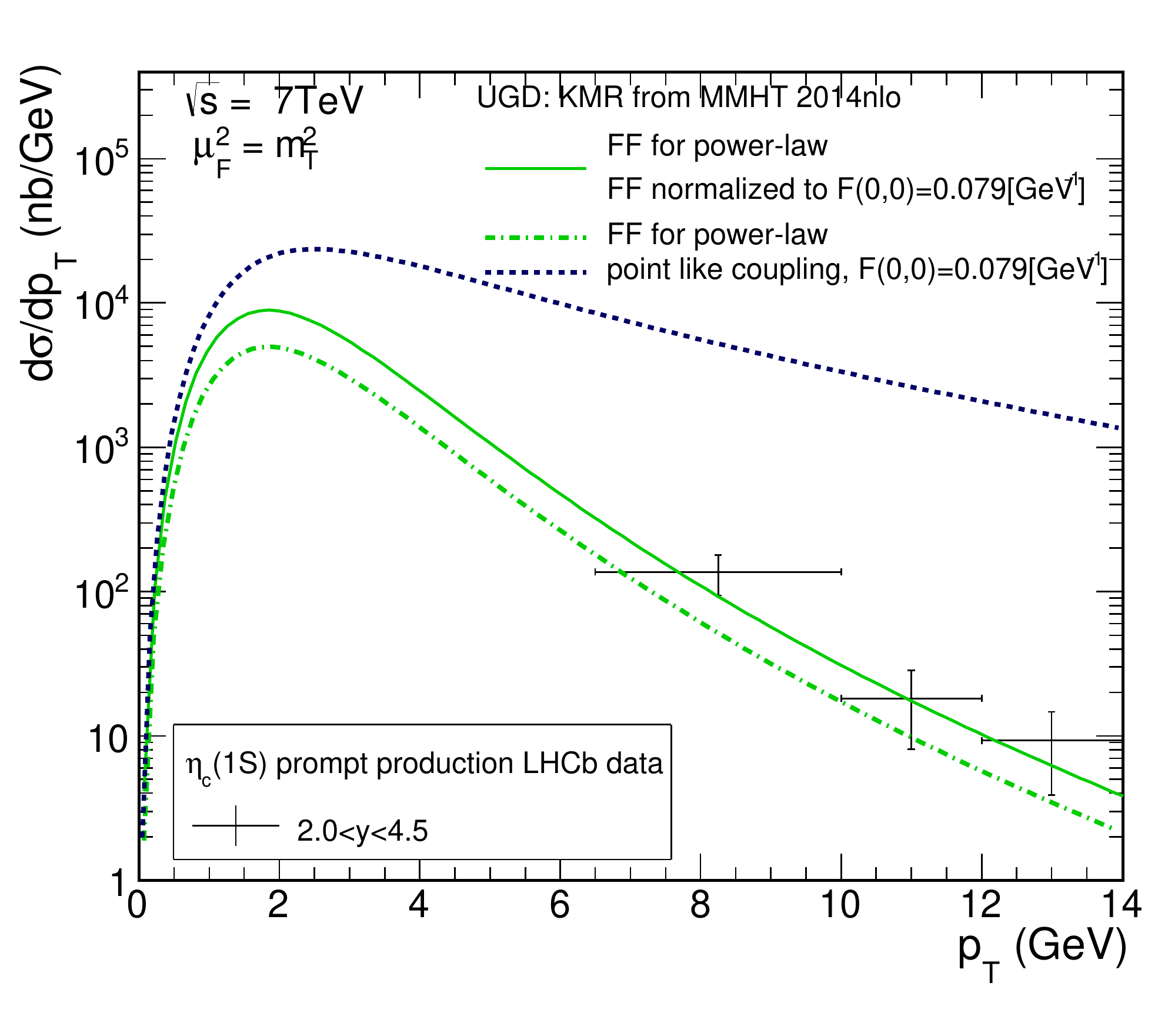}
        b) \includegraphics[width=0.4\textwidth]{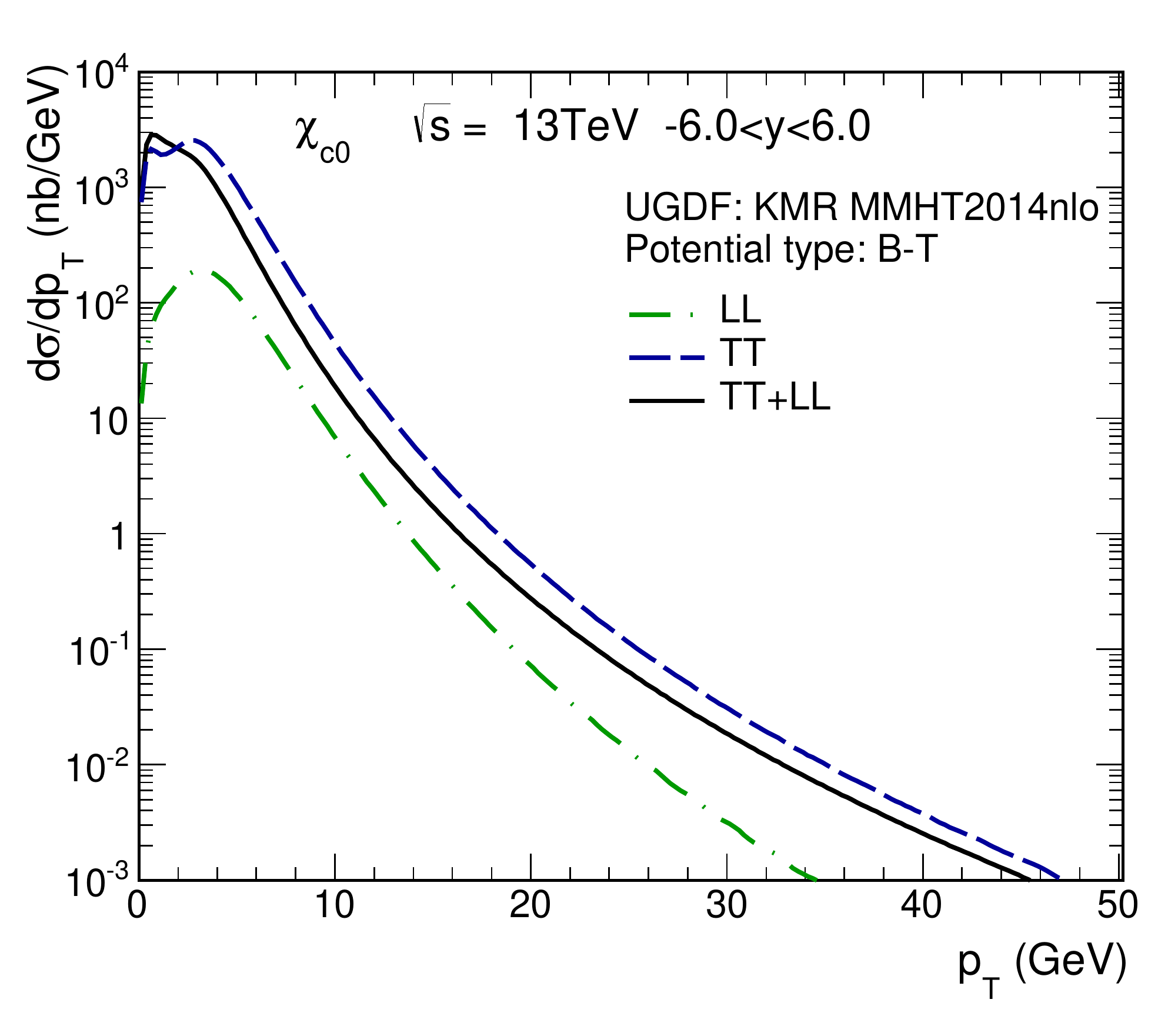}
        \caption{Panel a): Differential cross section for $\eta_c$ production in pp collisions at $\sqrt{s} = 7 \, \rm{TeV}$. The dashed curve shows the result obtained neglecting the off-shellness of gluons in the hard matrix element. Panel b): results for $\chi_{c0}$ production. We show the helicity decomposition of the cross section encoded in two independent form factors.}
        \label{fig:effect_of_ff}
\end{figure}

The amplitude for the $\gamma^*(Q_1^2) \gamma^*(Q_2^2) \to {\rm Meson}$ process 
is then obtained as a convolution of the $\gamma^* \gamma^* \to Q \bar Q$ amplitude with the light-front WF of the meson:
\begin{eqnarray}
{\cal M}_{\mu \nu} = \, \int {dz d^2\bk \over z(1-z) 16 \pi^3} 
\sum_{\lambda, \bar \lambda} 
\Psi^*_{\lambda \bar \lambda}(z,\bk) 
{\cal M}_{\mu \nu}^{\lambda \bar \lambda} (\gamma^* \gamma^* \to Q_\lambda(zP_+,\bp_Q) \bar Q_{\bar \lambda} ((1-z)P_+,\bp_{\bar Q}) \, . 
\label{eq:ampli}
\end{eqnarray}
The amplitude of Eq.\ref{eq:ampli} has a standard decomposition into invariant (transition-) form factors. For the pseudoscalar meson there exists one form factor $F(Q_1^2,Q_2^2)$, while for the scalar meson we have two form factors, $F_{TT}(Q_1^2,Q_2^2)$ and $F_{LL}(Q_1^2,Q_2^2)$, for the fusion of  photons with transverse and longitudinal polarizations in the $\gamma \gamma$-frame respectively. For all the form factors representations in terms of the light-front WFs have been derived in \cite{Babiarz:2019sfa,Babiarz:2019mag,Babiarz:2020jkh}. In Fig.\ref{fig:formfactors}a) we compare the results of our calulations for the $\eta_c$ transition form factor for one virtual photon, $F(Q^2,0)$ with data taken by the BaBar collaboration. The results for WFs from different potential models are shown. 
In Fig.~\ref{fig:formfactors}b) we show $Q^2 F(Q^2,0)$, which asymptotically is known to approach a constant (indicated by the band for different WFs). Evidently, in the practically relevant range of $Q^2$, we are far away from the asymptotic region.
The results for the transition form factors can be immediately used also in hadroproduction, say in $pp$ collisions. One starts from the $k_T$-factorization formula for the inclusive meson production cross section: 
	\begin{eqnarray}
	&&{d \sigma \over dy d^2\bp} = \int {d^2 \bq_1 \over \pi \bq_1^2} 
	{\cal{F}}(x_1,\bq_1^2) \int {d^2 \bq_2 \over \pi \bq_2^2}  {\cal{F}}(x_2,\bq_2^2)
	\, \delta^{(2)} (\bq_1 + \bq_2 - \bp ) \, {\pi \over (x_1 x_2 s)^2} \overline{|{\cal{M}}|^2} \, . 
	\end{eqnarray}
Here ${\cal{F}}(x,\bq^2)$ is the unintegrated gluon distribution (UGD) for the gluon with transverse momentum $\bq^2$ and spacelike virtuality $q^2 = - \bq^2$.
The off--shell matrix element 
\begin{eqnarray}
	{\cal{M}}^{ab} = {q_{1 \perp}^\mu q_{2\perp}^\nu \over |\bq_1| |\bq_2|}{\cal{M}}^{ab}_{\mu \nu}  = {q_{1+} q_{2-} \over |\bq_1| |\bq_2|} n^+_\mu n^-_\nu {\cal{M}}^{ab}_{\mu \nu} = {x_1 x_2 s \over 2 |\bq_1| |\bq_2| } n^+_\mu n^-_\nu {\cal{M}}^{ab}_{\mu \nu}  \, , 
	\end{eqnarray}
is, up to a trivial color factor, identical to the one of Eq.\ref{eq:ampli} and thus inherits the same off-shell form factors.
Results for inclusive $\eta_c$ production from Ref.\cite{Babiarz:2019mag} are shown in Fig.~\ref{fig:dsig_dpt_etac1S}a)-c), for a choice of UGDs. While there is a sizeable dependence on the UGD at lower $p_T=|\bp|$, at large $p_T$ a reasonable description of LHCb data at $\sqrt{s} = 7$ and 8 GeV is achieved. The data at $\sqrt{s} = 13 \, \rm{TeV}$ show a somewhat harder $p_T$-tail than our calculations. The role of the off-shell formfactor is crucial. This can be seen in Fig.~\ref{fig:effect_of_ff}a), where the result neglecting gluon off-shellness dramatically overshoots the data. Another feature of the off-shell approach is the polarization of gluons. Indeed the unintegrated gluon density 
is in fact better understood in terms of a whole polarization density matrix. In the on-shell limit, gluons are linearly polarized, while in the  general off-shell situation also longitudinal gluons are present. These have an effect in the case of the scalar particle. In Fig.~\ref{fig:effect_of_ff}b) we show the $p_T$--dependent cross section for $\chi_{c0}$
production in $pp$ collisions. The contributions related to $TT$ and $LL$ form factors are shown. We note that longitudinal gluons can affect the shape of the distribution in the peak region. Notice that an interference between $TT$ and $LL$ contributions is contained in the full result.

\section{Conclusion}
In summary, the color-singlet $k_T$-factorization approach leads to good description of LHCb data for inclusive $pp \to \eta_c$ for $\sqrt{s} = 7,8$ TeV, it fares  somewhat worse for large $p_T$ at $\sqrt{s} = 13$ TeV. Some room for a possible color octet contribution is left. In the LHCb kinematics very small $x$ are probed. Calculations show, that in the asymmetric kinematics with one large-$x$ and one small-$x$ parton, it is the  small-$x$ gluon transfers bulk of $p_T$.
Despite the sensitivity to small $x$,  no sign of gluon saturation is observed, the possible effects of the latter are concentrated at lower $p_T$. 
The $g^* g^* \eta_c, g^* g^* \chi_c$ form factors are related to $\gamma^* \gamma^*$ transition
form factors. Here, for the $\eta_c$ we can check our results against the
BaBar data, where we get in general a good description of the $Q^2$ dependence. 
For the case of the scalar particles $\chi_{c0}, \chi_{b0}$ a gluon density matrix of transverse and longitudinal polarizations are probed.
Predictions for $\chi_{c0},\chi_{b0}$ at LHC are available, but these mesons are difficult to measure.      

\paragraph{Funding information}
The  work  reported here was  partially  supported  by  the  Polish  National  Science  Center (NCN) grant  UMO-2018/31/B/ST2/03537, and by
the Center for Innovation and Transfer of Natural Sciences and  Engineering  Knowledge  in  Rzesz\'ow. I.B. was partially supported by  the Polish  National Agency  for  Academic  Exchange (NAWA) under  Contract  No. PPN/IWA/2018/1/00031/U/0001.
Part of this work has been performed in the framework of COST Action CA15213 Theory of hot matter and relativistic heavy-ion collisions (THOR).



\bibliography{SciPost_Example_BiBTeX_File.bib}

\nolinenumbers

\end{document}